\documentstyle[twocolumn,floats,aps]{revtex}
\newcommand{\be}{\begin{equation}}
\newcommand{\ee}{\end{equation}}
\newcommand{\bea}{\begin{eqnarray}}
\newcommand{\eea}{\end{eqnarray}}

\begin{document}
\draft 
\twocolumn[\hsize\textwidth%
\columnwidth\hsize\csname@twocolumnfalse\endcsname
\title{
{Circulation--Strain Sum Rule in Stochastic Magnetohydrodynamics}}
\author{ L. Moriconi$^{a}$ and F.A.S. Nobre$^{b,c}$}
\address{$a)$ Instituto de F\'\i sica, Universidade Federal do Rio de Janeiro \\C.P.
68528, Rio de Janeiro, RJ --- 21945-970, Brasil}
\address{$b)$ Centro Brasileiro de Pesquisas F\'\i sicas, Rua Xavier Sigaud, 150\\
Rio de Janeiro, RJ --- 22290-180, Brasil}
\address{$c)$ Universidade Regional do Cariri, rua Cel. Ant\^onio Luis, 1160\\ Crato, CE -- 63100-000, Brasil}
\maketitle
\begin{abstract}  
We study probability density functions (pdfs) of the circulation of velocity and magnetic fields in magnetohydrodynamics, computed for a circular contour within inertial range scales. The analysis is based on the instanton method as adapted to the Martin-Siggia-Rose field theory formalism. While in the viscous limit the expected gaussian behaviour of fluctuations is indeed verified, the case of vanishing viscosity is not suitable of a direct saddle-point treatment. To study the latter limit, we take into account fluctuations around quasi-static background fields, which allows us to derive a sum rule relating pdfs of the circulation observables and the rate of the strain tensor. A simple inspection of the sum rule definition leads straightforwardly to the algebraic decay $\rho(\Gamma) \sim  1/ \Gamma^2$ at the circulation pdf tails.
\end{abstract}
\vskip1pc] 
\narrowtext
\section{Introduction}
Probability distribution functions (pdfs) of quantities which ``measure" strong fluctuations in a turbulent fluid, like velocity differences in Burgers model \cite{burgers,yakhot,poly,gura} or the circulation in three-dimensional incompressible flows \cite{mig,cao,mori} are a very promising tool in the study of the intermittency phenomenon. While velocity differences are related to the existence of shock waves in the former case, the choice of circulation as an observable worth to be investigated is motivated by the appealing picture of turbulence in terms of the complex evolution of vorticity filaments, suggested for the first time through direct numerical simulations of the Navier-Stokes equations \cite{she-hoso}.

An early theoretical analysis of the circulation statistics by Migdal \cite{mig}, suggested, with the help of loop equation methods, and on the grounds of the central limit theorem, that the pdf of the circulation, evaluated for a closed contour contained in the inertial range, should display gaussian tails. This was not confirmed in direct numerical simulations, where pdfs were seen to exhibit a stretched exponential decay \cite{cao}. A further study of the problem by Takakura and one of us \cite{mori}, through the instanton approach \cite{falko}, was able to reproduce observed results as well as to establish predictions concerning the role of parity breaking external conditions, which presently stand open for experimental verification.

Our aim here is to address similar questions in the problem of turbulent
magnetohydrodynamics, motivated by the fact that a deeper understanding of the subject has been
in order for some time, in face of the impressive amount of data recorded in
astrophysical observations, as sunspots and the dynamo effect \cite{battaner}.
More specifically, we will study the statistics of circulation of velocity
and magnetic fields in the laminar and turbulent regimes, by means of the Martin-Siggia-Rose
formalism (MSR) \cite{msr}. The basic technical framework is provided by the
saddle-point method, where instantons and fluctuations around them are assumed to
yield, respectively, leading and subleading contributions to the behaviour of pdfs tails.
 
This paper is organized as follows. In sec. II, the basic elements of the formalism are set. We define the stochastic fluid equations in terms of the Elsasser variables (linear combinations of velocity and magnetic fields) from which the MSR action and the associated saddle-point equations are obtained. In Sec. III, as a concrete introduction to the instanton technique, we discuss the simpler problem of fluctuations in the viscous limit of the magnetohydrodynamical equations, proving circulation is governed in this case by exact gaussian statistics, as expected on physical grounds. In sec. IV we consider the limit of vanishing viscosity. It is known that a direct application of the saddle-point method is fated to fail here, as implied from simple dimensional analysis \cite{mori,falko}. We circumvent this problem through the strategy devised in ref. \cite{mori}, which consists of ``breaking" the path-integration measure into fast and slow degrees of freedom, related, respectively, to the vorticity and the rate of the strain tensor. The characterization of fast and slow variables is the basic physical assumption of our method, as motivated by the numerical experiments \cite{cao} performed in the case of pure turbulence. The outcome of our computations will be just a sum rule relating the statistics of circulation observables and the rate of the strain tensor. We find then asymptotic expressions for the tails of the circulation pdfs, given by the algebraic decay $1 / \Gamma^2$ (the same for velocity and magnetic circulation), a result that indicates the existence of strong intermittency effects in turbulent magnetohydrodynamics. In sec. V, we summarize our findings and discuss directions of further research.

\section{MSR formalism and the saddle-point equations}
Consider a three-dimensional fluid described by velocity and magnetic fields,
$v_\alpha( \vec x,t)$ and $b_\alpha (\vec x,t)$, respectively, which is stirred by
large scale stochastic forces. It is possible to choose physical units so that velocity and magnetic
fields have the same dimensions, and magnetohydrodynamical equations are written in terms
of Elsasser variables $z_\alpha^\pm = (v_\alpha \pm
b_\alpha)$ \cite{elsa}, as
\bea
&& \partial_t  z_\alpha^\pm + z_\beta^\mp \partial_\beta z_\alpha^\pm
=-\partial_\alpha P^\pm + \nu_+ \partial^2 z_\alpha^\pm +
\nu_- \partial^2 z_\alpha^\mp + f_\alpha^\pm  \ , \  \nonumber \\
&& \partial_\alpha z_\alpha^\pm = 0 \ , \  \label{msr1}
\eea
where $\nu_\pm = (\nu \pm \nu_b)/2$ are the viscosity parameters, which account for
dissipative effects, and $f_\alpha^\pm = (f_\alpha \pm g_\alpha)$ are gaussian
random external forces, defined through the correlators
\bea 
&&<f^\pm_\alpha (\vec x, t)> = <f_\alpha^\pm (\vec x, t) f_\beta^\mp (\vec
x',t') > = 0 \ , \ \nonumber \\ 
&&<f_\alpha^\pm (\vec x, t)
f_\beta^\pm (\vec x',t') > = \nonumber \\
&&=2 \delta(t-t')[D_{\alpha \beta}^{(1)} (\vec x -\vec x') \pm D_{\alpha \beta}^{(2)}
(\vec x - \vec
x') ]  \nonumber \\
&& \equiv \delta( t - t') D^{\pm}_{\alpha \beta}(\vec x - \vec x') \ , \  \label{msr2} 
\eea
with
\bea
D_{\alpha \beta}^{(1)} (\vec x - \vec x') &&=
<f_\alpha (\vec x, t) f_\beta (\vec x',t') >  \nonumber \\
&&= <g_\alpha (\vec x, t) g_\beta (\vec x',t') >  \nonumber \\
&&=D_1 \exp \left ( - |\vec x - \vec x'|^{2p} / L_1^{2p} \right ) \delta_{\alpha \beta}  \ ,
\ \nonumber \\
D_{\alpha \beta}^{(2)} (\vec x - \vec x') &&=
<f_\alpha (\vec x, t) g_\beta (\vec x',t') >  \nonumber \\
&&=D_2 \exp \left ( -|\vec x - \vec x'|^{2p} / L_2^{2p} \right ) \delta_{\alpha \beta}  \ .
\ \label{msr3} 
\eea
Above, $L_1$ and $L_2$ are the length scales where energy pumping mechanisms are assumed to occur, and $p > 0$ parametrizes the spatial decay of the force-force correlation functions. We note that an extension of the subsequent analysis to alternative definitions of the stochastic stirring terms yields no further computational difficulties.

It is important to determine functionals of the velocity and
magnetic fields that may carry useful information on intermittent behaviour,
to be revealed from the tails of probability distribution functions. 
A reasonable choice, as discussed in the introduction, is the
circulation, defined in terms of Elsasser
variables as
\be
\Gamma^\pm = \oint_c \vec z^\pm \cdot d \vec x \ . \ \label{msr4}
\ee
The above integral is evaluated at time $t=0$, assuming the fluid evolution
started at $t= - \infty$. The integration contour $c$ is taken to be the 
circumference $x^2 + y^2 = R^2$, with
$z=0$, oriented in the counterclockwise direction. $R$ is a length
contained in the inertial range, that is, $\eta \ll R \ll L_1,L_2$, where $\eta$
gives the microscopic length scale, associated to dominant viscous effects. A further
physical motivation to study the statistics of (\ref{msr4}) comes from
phenomena usually
observed in the astrophysical context: due to Amp\`ere-Maxwell's law,
the formation of magnetic tubes is related to strong fluctuations of
velocity circulation.
On the other hand, by the same token, any circulation of the magnetic
field leads to the existence of conduction/displacement currents, and possible
particle ``jets", which cross a surface bounded by the contour $c$.

The joint pdf for the circulation variables may be defined as
\bea
\rho (\Gamma^+,\Gamma^-) &&=({1 \over {2 \pi}})^2
\int_{- \infty}^{\infty} d \lambda^+
\int_{- \infty}^{\infty} d \lambda^- \nonumber \\
&& \times \exp(i \lambda^+ \Gamma^+ +i \lambda^- \Gamma^-) Z(\lambda^+,\lambda^-) \ ,
\ \label{msr5}
\eea
where the characteristic function is given by
\bea
&&Z(\lambda^+,\lambda^-) = \nonumber \\
&&= \langle \exp [ - i \lambda^+  \oint_c \vec z^+
\cdot d \vec x -i \lambda^- + \oint_c \vec z^- \cdot d \vec x ] \rangle \ . \
\label{msr6}
\eea

In the following computations, we will consider the analytical mapping $\lambda^\pm \rightarrow i \lambda^\pm$, so that in the large $\lambda^\pm$ limit the characteristic function will essentially pick up contributions from large fluctuations of the circulation (it is not difficult to prove this is so for pdfs which decay faster than any simple exponential). 

The stochastic partial differential equations (\ref{msr1}) may be studied as a
field theoretical problem. The MSR formalism \cite{msr} allows us to write the 
path integral expression
\be
Z(i\lambda^+,i\lambda^-)= \int D \hat z^\pm D Dz^\pm
DP^\pm DQ^\pm \exp(-S) \ , \ \label{msr7}
\ee
where the MSR action is 
\bea
S =&&-i \int d^3 \vec x dt [ \hat z_\alpha^+ (\partial_t z_\alpha^+
+ z_\beta^-  \partial_\beta  z_\alpha^+ -\nu_+ \partial^2 z_\alpha^-\nu_- \partial^2
z_\alpha^- \nonumber \\
&&+\partial_\alpha P^+) + \hat z_\alpha^- (\partial_t z_\alpha^- + z_\beta^+
\partial_\beta  z_\alpha^- -\nu_+ \partial^2 z_\alpha^-
-\nu_- \partial^2 z_\alpha^+ \nonumber \\
&&+\partial_\alpha P^-) + Q^+\partial_\alpha z_\alpha^+ +Q^+\partial_\alpha
z_\alpha^+ ] \nonumber \\
&&+{1 \over2} \int dt d^3 \vec x d^3 \vec x' \hat z_\alpha^+(\vec x,t)
D_{\alpha \beta}^+(\vec x - \vec x') \hat z_\beta^+ (\vec x', t) \ \nonumber \\
&&+{1 \over 2} \int dt d^3 \vec x d^3 \vec x' \hat z_\alpha^-(\vec x,t)
D_{\alpha \beta}^-(\vec x - \vec x') \hat z_\beta^- (\vec x', t) \nonumber \\
&&- \lambda^+ \Gamma^+ - \lambda^- \Gamma^- \ . \ \label{msr8}
\eea
A ``telegraphic" proof of the above formulation is based on the fact that
\bea
&&\exp[-{1 \over2} \int dt d^3 \vec x d^3 \vec x' \hat z_\alpha^\pm (\vec x,t)
D_{\alpha \beta}^\pm (\vec x - \vec x') \hat z_\beta^\pm (\vec x', t)]
\ \nonumber \\
&&=\langle \exp[i \int dt d^3 \vec x \hat z_\alpha^\pm (\vec x,t) f_\alpha^\pm
(\vec x,t)] \rangle_{f^\pm} \ , \ \label{msr}
\eea
where $\langle$(...)$\rangle_f$ stands for the average taken in the 
ensemble generated from different realizations of the stochastic forces.
Integration over the ``response fields" $\hat z_\alpha^\pm$ yields, thus, the
fluid equations (\ref{msr1}).
The role of the $P^\pm$ and $Q^\pm$ is just to assure that $\partial_\alpha z_\alpha^\pm = \partial_\alpha \hat z_\alpha^\pm =0$.
In the continuum time formulation, it is necessary to introduce a jacobian
term in (\ref{msr7}). However, if time is regarded as a discrete variable
(as we implicitely may do) there is no need for a jacobian (it is unity),
usually related to anticommuting Grassmann fields \cite{zinn}.

The MSR path-integral formulation may be used to recover the Wyld perturbative diagrammatic expansion \cite{wyld} of velocity correlation functions. Since the diagrammatic expansion is developed in powers of the convection terms, the perturbative MSR-Wyld approach has been severely criticized along the years for not taking into account singular configurations of the velocity field, which are of fundamental importance in turbulence. Nevertheless, one advantage of the MSR formalism is to address non-perturbative issues, from the knowledge of specific configurations of the flow that give relevant contributions to the path integral expression for $Z(i\lambda^+,i\lambda^-)$. This is precisely the task of the saddle-point method, meaningful in the limit of large $\lambda^+$ or $\lambda^-$. 

From (\ref{msr8}) we get the saddle-point equations
\bea
\partial_\alpha z_\alpha^\pm &&=0 \ , \ \nonumber \\
\partial_\alpha \hat z_\alpha^\pm &&=0 \ , \ \nonumber \\
 \partial_t  z_\alpha^\pm &&+ z_\beta^\mp \partial_\beta
z_\alpha^\pm - \nu_+ \partial^2 z_\alpha^\pm - \nu_- \partial^2 z_\alpha^\mp
+ \partial_\alpha P^\pm
= \nonumber \\
&&= - i\int d^3 \vec x' D_{\alpha \beta}^\pm (|\vec x -\vec x'|)
\hat z_\beta^\pm (\vec x',t) \ , \ \nonumber \\
\partial_t \hat z_\alpha^\pm &&- z_\beta^\mp \partial_\beta
\hat z_\alpha^\pm + \hat z_\alpha^\pm \partial_\beta z_\beta^\mp
+ \hat z_\beta^\mp \partial_\alpha z_\beta^\mp + \nu_+ \partial^2 \hat
z_\alpha^\pm \nonumber \\
+ \nu_- && \partial^2 \hat  z_\alpha^\mp + \partial_\alpha Q^\pm
= -i \lambda^\pm  {{\delta \Gamma^\pm} \over {\delta z_\alpha^\pm}} \ , \ \label{msr9}
\eea
with
\be
{{\delta \Gamma^\pm} \over {\delta z_\alpha^\pm}} =
\epsilon_{3 \beta \alpha} {x_\beta \over r_\perp} \delta (r_\perp-R)
\delta(z) \delta(t) \ , \
\label{msr10}
\ee
where $r_\perp =(x^2+y^2)^{1/2}$.

A major source of difficulty here is a hampering ``no-go" result that holds in the limit of vanishing viscosity: the 
saddle-point action computed from solutions of (\ref{msr9}) will necessarily depend on $\lambda^\pm$ in a way incompatible with parity symmetry. In fact, a simple check shows that saddle-point equations are invariant under a set of scaling transformations \cite{falko,mori}, which imply that the saddle-point action has the general form $S^{(0)} \sim \lambda^{3/2}$, if one takes $\lambda^\pm \equiv \lambda$ and $\nu \rightarrow 0$. This dependence on $\lambda$ is exactly the one found in Burgers turbulence for the statistics of velocity differences \cite{poly,gura}, which we do not expect to reproduce, even qualitatively, the parity symmetric pdfs of circulation in three-dimensions. In order to find physically meaningful results, a solution of this problem was advanced in ref. \cite{mori}, resorting on an alternative formulation of the MSR path-integral, where the rate of the strain tensor is used to parametrize an infinite family of saddle-point configurations. We will come back to this point in sec. IV.

\section{Viscous limit of Circulation pdfs}
It is interesting to study the viscous limit, where the convection terms are neglected in the fluid equations (in this case the above no-go theorem does not apply). Let us set $p=1$ in (\ref{msr3}). As a result, we get an instructive example where the circulation pdf may be exactly found. The saddle-point equations (\ref{msr9}) are now replaced by
\bea
\partial_t  z_\alpha^\pm &&+ \nu_+ \partial^2 z_\alpha^\pm
- \nu_- \partial^2 z_\alpha^\mp
=  \nonumber \\
&&=-i\int d^3 \vec x' D_{\alpha \beta}^\pm (|\vec x -\vec x'|)
\hat z_\beta^\pm (\vec x',t) \ , \ \nonumber \\
\partial_t \hat z_\alpha^\pm &&+ \nu_+ \partial^2 \hat
z_\alpha^\pm + \nu_- \partial^2 \hat z_\alpha^\mp
= -i \lambda^\pm {{\delta \Gamma^\pm} \over {\delta z_\alpha^\pm}} \ . \ \label{vl1}
\eea

Taking eqs. (\ref{msr8}) and (\ref{vl1}), it is possible, with the help of some partial integrations, to recast the saddle-point action in the simpler form
\be
S( \lambda,\lambda_b ) = - {\lambda} \oint_c \vec v \cdot d \vec x
- {\lambda_b} \oint_c \vec b \cdot d \vec x \ , \ \label{vl2}
\ee
where we used $\lambda^\pm = (\lambda \pm \lambda_b)/2$. It is also convenient to define
$\hat v_\alpha$ and  $\hat b_\alpha$ through $\hat z^\pm_\alpha = (\hat v_\alpha \pm \hat b_\alpha)/2$.
All we need to do, therefore, is to find solutions for the velocity and magnetic fields at $z=0$ and $t=0$, which we denote by $v_\alpha (\vec x_\perp,0)$ and $b_\alpha (\vec x_\perp,0)$, respectively, and to replace them in equation (\ref{vl2}).

We may rewrite equations (\ref{vl1}) as,
\bea
(\partial_t   - \nu \partial^2 )v_\alpha 
=&&-i \int d^3 \vec x' D_{\alpha \beta}^{(1)}(|\vec x -\vec x'|)
\hat v_\beta (\vec x',t) \nonumber \\
&&-i\int d^3 \vec x' D_{\alpha \beta}^{(2)}
(|\vec x -\vec x'|) \hat b_\beta (\vec x',t) \ , \ \nonumber \\
(\partial_t + \nu \partial^2 )\hat v_\alpha
= &&-i\lambda \epsilon_{3 \beta \alpha} {x_\beta \over r_\perp}
\delta (r_\perp-R) \delta(z) \delta(t) \ , \ \nonumber \\
( \partial_t - \nu_b \partial^2 ) b_\alpha 
= &&-i\int d^3 \vec x' D_{\alpha \beta}^{(1)}(|\vec x -\vec x'|)
\hat b_\beta (\vec x',t) \nonumber \\
&&-i \int d^3 \vec x' D_{\alpha \beta}^{(2)}
(|\vec x -\vec x'|) \hat v_\beta (\vec x',t)\ , \ \nonumber \\
(\partial_t + \nu_b \partial^2 ) \hat b_\alpha
=&&-i \lambda_b \epsilon_{3 \beta \alpha} {x_\beta \over r_\perp}
\delta (r_\perp-R) \delta(z) \delta(t) \ , \ \label{vl3}
\eea
Applying $( \partial_t + \nu \partial^2 )$ on the first equation listed in (\ref{vl3}) and using also the equations for $\hat v_\alpha$ and $\hat b_\alpha$, we will have
\be
[ \partial_t^2  - \nu^2 (\partial^2)^2] v_\alpha (\vec x,t) =
- F_\alpha (\vec x,t) \ , \
\label{vl4}
\ee
where
\bea
&&F_\alpha (\vec x,t) = \nonumber \\
&&=- \lambda \int d^3 \vec x' D_{ \alpha \beta}^{(1)}
(|\vec x - \vec x'|) \epsilon_{3 \gamma \beta} { x_\gamma' \over r_\perp'}
\delta (r_\perp' - R) \delta (z') \delta(t) \nonumber \\
&&-({{ \partial_t + \nu \partial^2 } \over { \partial_t + \nu_b \partial^2}})
\lambda_b \int d^3 \vec x' D_{ \alpha \beta}^{(2)}
(|\vec x - \vec x'|) \epsilon_{3 \gamma \beta} { x_\gamma' \over r_\perp'}
\delta (r_\perp' - R) \nonumber \\ 
&& \times \delta (z') \delta(t) \simeq { {D_1 \lambda 2 \pi R^2} \over L_1^2}
\epsilon_{3 \beta \alpha}
x_\beta \exp ( - {{\vec x^2} \over {L^2}}) \nonumber \\
&& + ({{ \partial_t  + \nu \partial^2 } \over { \partial_t  + \nu_b \partial^2}})
{ {D_2 \lambda_b 2 \pi R^2} \over L_1^2} \epsilon_{3 \beta \alpha}
x_\beta \exp ( - {{\vec x^2} \over L_2^2}) \ . \ \label{vl5}
\eea
In Fourier space, equation (\ref{vl5}) becomes
\be
( \omega^2 + \nu^2 k^4 ) \tilde v_\alpha (\vec k, \omega) =
\tilde F_\alpha (\vec k) \ . \
\label{vl6}
\ee
We obtain, thus,
\bea
&&v_\alpha (\vec x,t)
= ({ 1 \over {2 \pi }})^3 \int d^3 \vec k d \omega
{{\exp( i \vec k \cdot \vec x +i \omega t )} \over {\omega^2 + \nu^2 k^4}} \nonumber \\
&&\times \left [ { {\tilde F_\alpha^{(1)} (\vec k)} \over {\vec k^2}}+
({{i\omega - \nu \vec k^2} \over {i\omega - \nu_b \vec k^2}}) 
\tilde F_\alpha^{(2)} (\vec k)
 \right ] \ , \
\label{vl7}
\eea
with
\bea
&&\tilde F_\alpha^{(1)} (\vec k) =
-i \epsilon_{3 \beta \alpha} k_\beta { {D_1 \lambda \pi^{ {1 \over 2}}
R^2} \over 4} \exp( - { {L_1^2 \vec k^2} \over 4} ) \ , \ \nonumber \\
&&\tilde F_\alpha^{(2)} (\vec k) =
-i \epsilon_{3 \beta \alpha} k_\beta { {D_2 \lambda_b \pi^{ {1 \over 2}}
R^2} \over 4} \exp( - { {L_2^2 \vec k^2} \over 4} ) \ . \
\label{vl8}
\eea
Since we are interested to know $v_\alpha (\vec x_\perp,0)$, it follows,
from (\ref{vl7}),
\bea
&&v_\alpha (\vec x_\perp,0)= \int d^3 \vec k
\exp( i \vec k_\perp \cdot \vec x_\perp )  \nonumber \\
&&\times  { 1 \over {\vec k^2}}
\left [ { {\tilde F_\alpha^{(1)} (\vec k) } \over {4 \pi \nu}} 
+ { {\tilde F_\alpha^{(2)} (\vec k)} \over {2 \pi (\nu + \nu_b)}} 
 \right ] \ . \
\label{vl9}
\eea
Substituting (\ref{vl8}) in (\ref{vl9}), we find
\bea
v_\alpha (\vec x_\perp,0)&=&{{\pi R^2} \over 3}
\left [{ {  D_1 \lambda } \over {2 \nu}}
+{{ D_2 \lambda_b } \over {(\nu + \nu_b)}} \right ]
\epsilon_{3 \beta \alpha} x_\beta  \ .
\label{vl10}
\eea
Similarly,
\bea
b_\alpha (\vec x_\perp,0)&=&
{{\pi R^2} \over 3}
\left [ { {  D_1 \lambda_b } \over {2 \nu_b}}
+{{ D_2 \lambda_b } \over {(\nu + \nu_b)}} \right ]
\epsilon_{3 \beta \alpha} x_\beta  \ . \
\label{vl11}
\eea
Thus, from (\ref{vl2}), (\ref{vl10}) and (\ref{vl11}), we obtain
the saddle-point action
\bea
S(\lambda,\lambda_b)=
-\lambda^2 \eta_1-\lambda_b^2 \eta_2-\lambda\lambda_b \eta_3
\ . \
\label{vl12}
\eea
where
\bea
\eta_1 = {{D_1\pi^2R^4} \over {3\nu}} \ , \
\eta_2 = {{D_1\pi^2R^4} \over {3\nu_b}} \ , \
\eta_3 = {{4D_2\pi^2R^4} \over {3(\nu+\nu_b)}} \ .
\label{vl13}
\eea
Performing now $\lambda \rightarrow -i \lambda$ and $\lambda_b \rightarrow -i \lambda_b$, to restore the original definition of these parameters, we get
\be
Z( \lambda,\lambda_b) \propto \exp (-\lambda^2 \eta_1-\lambda_b^2 \eta_2
-\lambda\lambda_b \eta_3)\ . \
\label{vl14}
\ee
Defining
\be
\Gamma = \oint_c \vec v \cdot d \vec x \ {\hbox{ , }} \
\Gamma_b = \oint \vec b \cdot d \vec x \ , \ \label{circs}
\ee
we obtain, from (\ref{msr6}), the gaussian pdf
\be
\rho (\Gamma,\Gamma_b) = ({1 \over {4 \pi^2  \zeta}})^{1 \over 2} \exp \left [ -{{ \eta_2 \Gamma^2}
\over {4 \zeta}}
-{{ \eta_1 \Gamma_b^2} \over {4  \zeta }} +
{{\eta_3 \Gamma \Gamma_b } \over {4  \zeta}} \right ] \ , \
\label{vl15}
\ee
with
\bea
\zeta =\eta_1 \eta_2- {\eta_3^2 \over 4} > 0 \ . \
\label{vl16}
\eea
This condition may be stated as
\be
\sqrt{ {\nu \over \nu_b}} + \sqrt{{\nu_b \over \nu}} > { {2 D_2} \over D_1}
\ . \ \label{vl17}
\ee
It follows from (\ref{msr2}) and (\ref{msr3}) that the above inequality is always satisfied (in fact, $D_1 > D_2$ while
the lhs of (\ref{vl17}) is $\geq 2$).
Considering, for instance, $D_2=0$, which gives $\eta_3=0$, we find
the factorized form of the circulation pdf,
\bea
\rho (\Gamma,\Gamma_b)=({1 \over {4 \pi^2 \eta_1 \eta_2}})^{1 \over 2}
\exp \left [ -{{\Gamma^2} \over {4 \eta_1}} - {{\Gamma_b^2} \over {4 \eta_2}} \right ]
\ . \
\label{vl18}
\eea
\section{Inviscid theory and the Circulation-strain sum rule}
Let us study now the inviscid theory, where $\nu, \nu_b \rightarrow 0$, a limit associated to the fully developed
turbulent regime. Since the radius $R$ of the contour $c$ is much smaller than $L_1$ and $L_2$, the scales where energy is injected into the system, we are allowed to consider linear expressions for the fields $z^\pm_\alpha$,
\be
z_\alpha^\pm (\vec x ,t) = \sigma_{\alpha \beta}^\pm (t) x_\beta \ , \ \label{vslin}
\ee
with the tensor of velocity derivatives satisfying to $\sum_\alpha \sigma_{\alpha
\alpha}^{\pm}=0$, due to the incompressibility constraint and the absence of magnetic monopoles.
To work in the order of approximation given by (\ref{vslin}) we consider a quadratic form for the
pressure fields $P^\pm$,
\be
P^\pm=A_{\alpha \beta}^\pm x_\alpha x_\beta \ , \
\label{vs1}
\ee 
so that the gradient terms $ \partial_\alpha P^\pm $ exactly cancel in (\ref{msr9}) any symmetric tensor acting on the
spatial coordinates, which would appear in the linear approximation. We are left
this way with equations which describe the time evolution of the antisymmetric part of $\sigma_{\alpha \beta}^\pm$ (related to the vorticity)
\bea
{d \over { dt}} \sigma^{\pm \bar s}_{\alpha \beta} &&= 
- (\sigma^{\pm s}
\sigma^{\mp \bar s} +  \sigma^{\pm \bar s} \sigma^{ \mp s})_{\alpha \beta} \nonumber \\
&&- i \int d^3 \vec x \partial_{[\alpha ,} D_{\beta ] \gamma}^\pm (|\vec x|) \hat z_\gamma^\pm (\vec x ,t) 
\ , \  \label{vs2}
\eea
where we have defined
\bea
&& \sigma^{\pm s}_{\alpha \beta} = {1 \over 2} ( \sigma_{\alpha
\beta}^\pm + \sigma_{\beta \alpha}^\pm) \ , \ \sigma^{\pm \bar s}_{\alpha \beta} = {1 \over 2} ( \sigma_{\alpha
\beta}^\pm - \sigma_{\beta \alpha}^\pm) \ , \ \nonumber \\
&& \partial_{[\alpha ,} D_{\beta ] \gamma}^\pm (|\vec x|)= {1
\over 2} \left ( \partial_\alpha D_{\beta  \gamma}^\pm (|\vec x|)-
\partial_\beta D_{\alpha \gamma}^\pm (|\vec x|) \right ) \ . \ \label{vs3}
\eea

Before proceeding, it is necessary to discuss the alternative definition of the saddle-point method that we will employ.
It has been suggested through numerical experiments of pure turbulence \cite{cao} that the rate of strain tensor, $\sigma^s$, does not fluctuate so strongly as the vorticity. As it was shown in ref. \cite{mori}, one may take advantage of this physical observation to overcome the no-go result commented in sec. II. We will assume that the characterization of symmetric and antisymmetric degrees of freedom as slow and fast fluctuating variables, respectively, holds also in the magnetohydrodynamical realm, where similar vorticity filamentary structures are as well observed. Of course, experiments will decide ultimately if this assumption is correct or not, but we provisionally regard it as a working hypothesis that allows us to establish meaningful predictions. Following ref. \cite{mori} we may define the MSR path-integral for the characteristic functional as
\bea
Z(i\lambda^+,i\lambda^-)&&=  \int D \sigma^{\pm s} \int D \hat z^\pm Dz^\pm \nonumber \\
&& \times DP^\pm
DQ^\pm D \tilde Q^\pm \exp(- \tilde S) \ , \
\label{vs5} 
\eea
with $\sigma^{\pm s}= \sigma^{\pm s} (x,y,t)$ and
\bea
\tilde S &&= S - {i \over 2} \int dx dy dt
\tilde Q_{\alpha \beta}^+(x,y,t) 
\cdot [ \partial_\alpha  z_\beta^+
|_{z=0} + \partial_\beta z_\alpha^+  |_{z=0} 
\nonumber \\
&&-2 \sigma^{+s}_{\alpha \beta}(x,y,t)] 
-{i \over 2} \int dx dy dt \tilde Q_{\alpha \beta}^-(x,y,t)
\cdot [ \partial_\alpha  z_\beta^- |_{z=0}
\nonumber \\
&&+ \partial_\beta z_\alpha^- |_{z=0} - 2 \sigma^{-s}_{\alpha
\beta}(x,y,t) ] \ . \ \label{vs6}
\eea
The functional form (\ref{vs5}) is obtained by putting
\be
1=\int D \sigma^{\pm s} D \tilde Q^\pm \exp(S - \tilde S) \ , \ \label{vs7}
\ee
in the integrand of (\ref{msr7}). The order of integrations is interchanged, to write the integral over $\sigma^{\pm s}_{\alpha \beta}$ as the last one to be performed. These mathematical steps are motivated by the fact that the rate of the strain tensor plays the role of a quasi-static background where vorticity fluctuations take place. The central idea is then to apply the saddle-point method to the action $\tilde S$, treating $\sigma^{\pm s}_{\alpha \beta}$ as external fixed fields. The saddle-point equations for $\hat z^\pm_\alpha$, in (\ref{msr9}), are replaced now by
\bea
&&\partial_t \hat z_\alpha^\pm - \hat z_\beta^\mp
\partial_\alpha z_\beta^\mp + z_\beta^\mp
\partial_\beta \hat z_\alpha^\pm + \hat z_\beta^\pm
\partial_\beta z_\alpha^\mp
+ \nu_+ \partial^2 \hat z_\alpha^\pm
\nonumber \\
&&+ \nu_- \partial^2 \hat z_\alpha^\mp 
+ \partial_\alpha Q^\pm +
\partial_\beta  ( \delta (z) \tilde Q_{\beta \alpha}^\pm) +i 
\lambda^\pm {{\delta \Gamma^\pm} \over {\delta z_\alpha^\pm}} = 0 \ , \ \label{vs8}
\eea
There are also two additional equations, associated to
variations of the fields $\tilde Q_{\alpha \beta}^\pm$,
\be
\partial_\alpha z_\beta^\pm  |_{z=0}
+\partial_\beta z_\alpha^\pm |_{z=0} - 2 \sigma^{\pm s}_{\alpha
\beta}(x,y,t) = 0  \ . \  \label{vs9}
\ee

In order to seek for solutions of the saddle-point eqs. (\ref{msr9}), note they are invariant under rotations around the $z$ axis. The most general form of an axisymmetric tensor of velocity derivatives is given by
\be
 \sigma^\pm (t) = \left[ \matrix{ a^\pm (t) & \omega^\pm (t) & 0\cr
 - \omega^\pm (t) & a^\pm (t) & 0\cr
  0 & 0 & -2a^\pm (t)} \right] \ . \ \label{vs4}
\ee
From now on, we will substitute $\sigma^{\pm s}_{\alpha \beta}$ appearing in the above relations by the axisymmetric expression (\ref{vs4}).
This approximation amounts to the replacement
\be
\int D \sigma^{\pm s}_{\alpha \beta} \rightarrow \int Da^\pm(t) \ , \ \label{vs10}
\ee
in the path-integral (\ref{vs5}). If one were able to integrate exactly the expression for $Z(\lambda^\pm)$, keeping
$a^\pm (t)$ fixed, the circulation pdf could be written as
\be
\rho(  \Gamma^\pm ) =\int D a^\pm
\bar \rho[  a^\pm ] \rho[ \Gamma^\pm  |  a^\pm ] 
\ , \ \label{sumrule}
\ee
where $\bar \rho[a^\pm ]$ is the probability density functional to get the axisymmetric rate of the strain tensors defined by $a^\pm(t)$, and 
$\rho[  \Gamma^\pm  |  a^\pm ]$ is the conditional pdf to get
$\Gamma^\pm$, in the background $a^\pm(t)$. A natural question is how to reproduce this kind of relation through the instanton approach. Is the saddle-point action leading to an approximation for $\bar \rho[a^\pm]$, $\rho[\Gamma^\pm | a^\pm]$, or both?. We do not know how to answer it a priori. We take a pragmatical point of view, where an answer is found only after concrete computations are performed.

Using (\ref{msr10}), (\ref{vslin}), (\ref{vs4}), and taking the limit of vanishing viscosity, we may write (\ref{vs8}) as
\bea
&&
\partial_t \hat z_\alpha^\pm - \sigma_{\beta \alpha}^\mp \hat
z_\beta^\mp + \sigma_{\beta \gamma}^\mp x_{\gamma}
\partial_\beta \hat z_\alpha^\pm + \partial_\alpha Q^\pm +  \partial_\beta
(\delta (z)  \tilde Q_{\beta \alpha}^\pm) = \nonumber
\\
&& =i\lambda^\pm \epsilon_{3 \alpha \beta} {x_\beta \over r_\perp}
\delta (r_\perp-R) \delta(z) \delta(t) \ . \ \label{vs11}
\eea
We have now a closed set of coupled equations given by (\ref{vs2}) and (\ref{vs11}).
A crucial observation is that the viscosity terms in (\ref{vs8}) have the opposite sign, if compared to the ones
appearing in the usual fluid equations. We have to impose, therefore, in order to avoid an unbounded growing of the fields
$\hat z_\alpha^\pm (\vec x,t)$, that $\hat z_\alpha^\pm =0$, for $t>0$. In this way, (\ref{vs11}) leads to the boundary condition
\be
\hat z_\alpha^\pm(\vec x, 0^-)= i\lambda^\pm \epsilon_{3 \beta
\alpha} {x_\beta \over r_\perp} \delta (r_\perp-R) \delta(z) \ . \
\label{vs12}
\ee 
Also, we require that $\hat  z_\alpha^\pm (\vec x
, t) \rightarrow 0$ as $t \rightarrow - \infty$. The equation for
$\hat z_\alpha^\pm (\vec x,t)$ may be solved through the ansatz
\be
\hat z_\alpha^\pm (\vec x, t) = \epsilon_{3 \beta \alpha} x_\beta
\delta (z) \sum_{n=0}^{\infty} c_n^\pm(t)
r_\perp^{n-1}\delta^{(n)} (r_\perp-R) \ , \ \label{vs13}
\ee
where
$\delta^{(n)}(r_\perp-R) = d^n \delta(r_\perp-R) / dr_\perp^n$.
The problem of finding $\hat z^\pm_\alpha$ is mapped into the computation of $c_n^\pm(t)$. The boundary condition (\ref{vs12}) reads now
\bea
&&c_0^\pm(0^-)=i \lambda^\pm \ , \ \nonumber
\\
&&c_n^\pm(0^-)=0,{\hbox{ for $n > 0$}} \ . \ \label{vs14}
\eea
We obtain, substituting (\ref{vs13}) in (\ref{vs11}),
\be
{d \over {dt}} C_n = [A-(2+n)B]C_n - BC_{n-1} \ , \ \label{vs15}
\ee
with $C_{-1} \equiv 0$ and
\bea
&&A (t) =\left [ \matrix{ 0 &a^-(t) \cr
  a^+(t)  & 0}\right] \ , \
B (t) =\left [ \matrix{ a^-(t) & 0 \cr
  0  & a^+(t)}\right] \ , \ \label{vs16}
\nonumber \\
&&C_n(t)= \left [ \begin{array}{c} c_n^+ \\ c_n^- \end{array}
\right ] \ . \  \label{vs17}
\eea
Furthermore, we get $Q=0$, and
\be
\tilde Q^\pm(r_\perp,t) = - 2\omega^\pm(t) \sum_{n=0}^{\infty} c_n^\pm(t)
\int_0^{ r_\perp} d \xi \xi^n \delta^{(n)} ( \xi- R )
 \ . \ \label{vs16.1}
\ee
An exact solution for $C_n(t)$ may be found in the case where
$[A,B] =0$. Matrices A and B commute only if
$a^+(t)=a^-(t) \equiv a(t)$, which means that
\bea
&&\partial_\alpha b_\beta + \partial_\beta b_\alpha = 0 \ , \ \nonumber \\
&&\partial_\alpha v_\beta + \partial_\beta v_\alpha =2a(t) \cdot 
(\delta_{\alpha \beta}
-3\delta_{3 \alpha} \delta_{3 \beta}) \ . \ \label{vs18}
\eea
We are concerned, thus, with fluctuations of the circulation variables 
in the presence of a vanishing rate of the strain tensor of the magnetic field.
Even though the need of computational feasibility compels us
to the study of the {\it{conditional pdf of circulation variables}} 
in flow configurations where (\ref{vs18}) is verified, this restriction is not unfortunate
by no means: we will be able, below, to find well-defined predictions related to intermittent 
fluctuations of the velocity or magnetic circulation. 

The exact solution of (\ref{vs15}) 
is given by 
\bea
&&C_n(t)={{i} \over {n!}} e^{ -\int_0^t dt'(2B-A)} \left( e^{
-\int_0^t dt' B} -1 \right)^n \tilde \lambda \ , \ \label{vs19}
\eea
where
\bea
\tilde \lambda = \left [ \begin{array}{c} \lambda^+ \\
\lambda^-
\end{array} \right ] \ . \ \label{vs20}
\eea
Taking into account (\ref{vs19}), the infinite series (\ref{vs13}) may be exactly summed up,
to yield, in terms of 
$\hat v_\alpha$ and $\hat b_\alpha$,
\bea
\hat v_\alpha(\vec x,t)&&= i\lambda \epsilon_{3 \beta \alpha}
{ x_\beta \over r_\perp} \delta (r_\perp-R  e^{ \int_0^t dt'
a(t')}) \delta(z) \ , \ 
\nonumber \\
\hat b_\alpha(\vec x,t)&&= i\lambda_b \epsilon_{3 \beta \alpha} {
x_\beta \over r_\perp}e^{- 2 \int_0^t dt' a(t')}\nonumber \\
&& \times \delta (r_\perp-R
e^{ \int_0^t dt' a(t')}) \delta(z) \ . \ \label{vs21}
\eea
To find the saddle-point action $\tilde S^{(0)}$, it is necessary to get $\omega^\pm(t)$.
Using (\ref{vs4}), we write eqs. (\ref{vs2}) as
\be
{d \over {dt}} \omega^\pm + a (\omega^+ + \omega^-) =-i \int d^3 \vec x \partial_{[ 1 ,} 
D_{ 2 ] \alpha}^\pm (|\vec x|) \hat z_\alpha^\pm (\vec x ,t) \ . \ \label{vs22}
\ee
Substituting the solutions for $\hat z_\alpha^\pm (\vec x, t)$ in (\ref{vs22}), we obtain, for $t < 0$, 
\bea
&&{d \over {dt}} \omega^\pm + a (\omega^+ + \omega^-) = \nonumber \\
&&=[l_1^\pm  + l_2^\pm 
e^{-2 \int_0^t dt' a(t')}]e^{2 p \int_0^t dt' a(t')} \ , \ \label{vs33}
\eea
where
\bea
&& l_1^\pm =-2 \pi p R^{2p} \lambda \left({{D_1} \over {L_1^{2p}}} \pm
{{D_2} \over {L_2^{2p}}}\right) \ , \ \nonumber \\
&&l_2^\pm= \mp 2 \pi p R^{2p}
\lambda_b
\left( {{D_1} \over {L_1^{2p}}} \pm {{D_2} \over {L_2^{2p}}} \right) \ . \ \label{vs34}
\eea
The idea now is to consider, in the context of a gradient expansion, the
effects of time independent configurations $a(t)=a$. Therefore, we replace the 
path-integration over arbitrary fields $a(t)$ by an ordinary integration
over $a$. Furthermore, it turns out that $\omega^\pm (t) \rightarrow 0$ as 
$t \rightarrow - \infty$ only if $a>0$ and $p>1$. Provided these conditions are satisfied,
we obtain
\bea
\omega^\pm (t) &=& { {(2p+1) l_1^\pm - l_1^\mp} \over {4 a p (p+1)}} \exp[2pat] \nonumber \\
&+&  { {(2p-1) l_2^\pm - l_2^\mp} \over {4 a p (p-1)}} \exp[2(p-1)at]  \ . \  \label{vs35}
\eea

The saddle-point action may be written as
\bea
\tilde S^{(0)} = &&- {1 \over 2} \int _{-\infty}^0 dt \int d^3 \vec x d^3 \vec x' \nonumber \\
&& \times [\hat
z^+_\alpha(\vec x , t)D_{\alpha \beta}^+ (\vec x - \vec x') \hat
z^+_\beta (\vec x' , t) \ \nonumber  \\
&&+ \hat
z^-_\alpha(\vec x , t)D_{\alpha \beta}^- (\vec x - \vec x') \hat
z^-_\beta (\vec x' , t)] \nonumber \\
&&- \lambda^+\oint_c \vec z^{+(0)} \cdot d \vec x -
\lambda^-\oint_c \vec z^{-(0)} \cdot d \vec x \ \label{vs36}
\eea
Using (\ref{vs21}) and (\ref{vs33}), a straightforward computation gives
a quadratic form in $\lambda$ and $\lambda_b$ for the action $\tilde S^{(0)}$. 
With the help of the notation introduced in the analysis of the viscous limit, equation 
(\ref{vl12}), we have now
\be
\eta_1 = { {D_1  p f(p)}  \over {L_1^{2p} (p+1)}} \ , \ 
\eta_2 = { {D_1  p f(p)}  \over {L_1^{2p} (p-1)}}  \ , \
\eta_3 = { {2 D_2 f(p)}  \over {L_2^{2p} }}   \ , \  \label{vs37}
\ee
where, in terms of gamma functions, 
\be
f(p) = { {\pi^2 R^{2(p+1)} } \over a } \left [ 
2 - { {\Gamma (2p+1)} \over { \Gamma(p+1) \Gamma(p+2)} } \right ] \ . \ \label{vs37.1}
\ee
The requirement of having a positive definite saddle-point action $\tilde S^{(0)}$ (after the analytical mapping $\lambda^\pm \rightarrow i \lambda^\pm$) leads to $f(p)>0$, that is $p<2$, and
\be
{1 \over p^2} > 1 - \left [ { {D_1 L_2^{2p}} \over {D_2 L_1^{2p}}} \right ]^2  \ . \ \label{vs37.2}
\ee
Therefore, our formalism is assumed to work in the inviscid theory for $1<p<2$, if the above condition is also verified (what happens, in particular, for $D_2=0$). It is interesting to observe that $\eta_2$ diverges when $p \rightarrow 1$, so that the pdf of the magnetic circulation becomes very broad in that limit. This fact may be physically interpreted as due to a progressive decoupling of magnetic and velocity fields as $p$ approaches unity, in flow realizations where the magnetic strain vanishes. The velocity field would then behave in the same fashion as in pure hydrodynamical turbulence, while the magnetic circulation would diffuse in a strong way, under the action of the external stochastic forces. On the other hand, for $p$ large enough, higher wavenumbers in the Fourier transform of the force-force correlation functions cannot be neglected, and the saddle-point method, as based on smooth instanton solutions, breaks down.

Taking the definitions for the $\eta$'s in (\ref{vs37}), the conditional pdf for the circulation variables,
$\rho (\Gamma, \Gamma_b | a )$, may be readily obtained from the expression previously defined in (\ref{vl15}). 
We may write now, recalling (\ref{sumrule}), the joint circulation pdf as
\be
\rho (\Gamma,\Gamma_b) = \int_0^\infty da \bar \rho (a) \rho (\Gamma, \Gamma_b | a) \ . \ \label{pdfvb}
\ee
This is a sum rule which holds for asymptotically large values of $\Gamma$ or $\Gamma_b$. It is important to keep in mind that the pdfs appearing in (\ref{pdfvb}) are defined under the condition that $\partial_\alpha b_\beta + \partial_\beta b_\alpha =0$. Also, it is natural to expect that the unknown function $\bar \rho(a)$ has a finite limit as $a \rightarrow 0$, so that we are led to the algebraic decay at the pdf tails
\bea
&& \rho(\Gamma, \Gamma_b) \sim \nonumber \\
&& \left [{ {D_1 p}  \over {L_1^{2p} (p-1)}} \Gamma^2 +
{ {D_1 p}  \over {L_1^{2p} (p+1)}} \Gamma_b^2 - { {2 D_2 }  \over {L_2^{2p}}} \Gamma \Gamma_b \right ]^{-3/2} \ , \ \label{pdftail}
\eea
a result which clearly signalizes the intermittent nature of circulation fluctuations in turbulent magnetohydrodynamics.
In particular, we may get from (\ref{pdftail}) the pdf tail for the magnetic circulation as 
\be
\rho(\Gamma_b) = \int d \Gamma \rho( \Gamma, \Gamma_b) \sim |\Gamma_b|^{-2} \ . \ \label{pdftail2}
\ee
An analogous behaviour follows for the velocity circulation pdf. Since the magnetohydrodynamic system is most of the time around the state where the magnetic strain vanishes (that is, the mean magnetic strain is zero), we conjecture that (\ref{pdftail2}) is a general result, holding beyond a specific selection of ensembles in the fully turbulent regime. It is worth mentioning that similar computations for the case of pure turbulence reveal ``less intermittent" fluctuations, i.e, the circulation pdf tail $\rho(\Gamma)\sim 1/| \Gamma |^3$. 

One could be puzzled by the fact that we have found algebraic decaying pdfs, exploring the large $\lambda^\pm$ limit, which, as commented before, is related to pdfs with fast exponential decay. However -- a crucial point in the analysis -- the instanton method becomes meaningful since it has been used to get the {\it{conditional}} pdf $\rho(\Gamma,\Gamma_b |a)$, which has an exact gaussian shape. We stress that the algebraic decay in eq. (\ref{pdftail}) follows from the integration over the velocity strain parameter $a$ in eq. (\ref{pdfvb}). Also, at this point it should be noted that the prediction of the stretched exponential tail for the circulation pdf in pure turbulence, as previously discussed in ref. \cite{mori}, is indeed likely to hold at an {\it{itermediate}} range where deviations from gaussian statistics appear, and not at the extreme asymptotic region. To describe the intermediate range -- the one so far observed in direct numerical simulations -- it is necessary to know in some detail the form of $\bar \rho(a)$, the strain pdf. 

\section{Conclusions}
We studied the problem of stochastic magnetohydrodynamics in the limits of large and small viscosity parameters, focusing on  circulation variables, which are expected to yield information on itermittent behaviour in the latter situation (turbulent regime). From the technical point of view, the computational framework was given by the application of the saddle-point method within the Martin-Siggia-Rose path-integral formalism. 

The instanton approach was straightforwardly applied to the viscous case, where we found a gaussian form for the circulation pdf, as it should be. Regarding the inviscid limit, when turbulence surely comes into play, our investigation is particularly devoted to the properties of pdf tails. The fundamental physical hypothesis is that for both velocity and magnetic fields, the rate of the strain tensor behaves as a quasi-static background where faster fluctuations of the antisymmetric part of the tensor of field derivatives (related to the vorticity) occur. This motivates a definition of the Martin-Siggia-Rose functional which explicitely takes into account fast and slow degrees of freedom. The saddle-point method is then applied to the action defined in terms of the fast variables. Such a strategy gives a way out of the problematic dimensional constraints imposed on the form of the saddle-point action in the original path-integral formalism. We were able in this way to obtain a 
sum rule relation for the statistics of the circulation variables, as well as a quantitative prediction for the algebraic decay of pdf tails.

Some mathematical simplifications were employed in the course of analysis, which basically fall into two classes: either based on the role of fast and slow variables or related to the manipulation of exact solutions of the saddle-point equations. Improvements on the latter aspect are likely to be the more interesting (and perhaps the more difficult to be attained). The relevant open question is then how to implement the instanton approach when exact saddle-point solutions are not available anymore, in order to compute joint pdfs of circulation variables in turbulent magnetohydrodynamics.

\acknowledgements
This work was partially supported by CAPES and CNPq.


\begin{references}
\bibitem{burgers}  J.M. Burgers, Adv. Appl. Mech. {\bf 1}, 171 (1948).
\bibitem{yakhot} A. Chekhlov and V.Yakhot, Phys. Rev. E {\bf 52}, 5681 (1995).
\bibitem{poly} A.M. Polyakov, Phys. Rev. E {\bf 52}, 6183 (1995).
\bibitem{gura} V. Gurarie and A. Migdal, Phys. Rev. E {\bf 54}, 4908 (1996).
\bibitem{mig} A. Migdal, Int. J. Mod. Phys. {\bf 9}, 1197 (1994).
\bibitem{cao} N. Cao, S. Chen and K.R. Sreenivasan, Phys. Rev. Lett. {\bf
76}, 616 (1996).
\bibitem{mori} L. Moriconi and F.I. Takakura, Phys. Rev. E {\bf 58}, 3187 (1998).
\bibitem{she-hoso} Z.S. She, E. Jackson and S.A. Orszag, Nature {\bf 344}, 226 (1990); I. Hosokawa and K. Yamamoto, J. Phys. Soc. Japan {\bf 59}, 401 (1990).
\bibitem{falko} G. Falkovich, I. Kolokolov, V. Lebedev and A. Migdal,
Phys. Rev. E {\bf 54}, 4896 (1996).
\bibitem{battaner} E. Battaner, {\it Astrophysical Fluid Dynamics},
Cambridge University Press (1996).
\bibitem{msr} P.C. Martin, E.D. Siggia and H.A. Rose, Phys. Rev. A {\bf 8},
423 (1973).
\bibitem{elsa} W.M. Elsasser, Phys. Rev. {\bf{79}}, 183 (1950).\
\bibitem{zinn} J. Zinn-Justin, {\it Field Theory and Critical Phenomena}, Claredon Press Oxford (1989).
\bibitem{wyld} H.W. Wyld, Ann. Phys. {\bf 14}, 143 (1961).
\end{references}
\end{document}